\documentclass[twocolumn]{revtex4-1}

\usepackage[utf8x]{inputenc}
\usepackage{amsmath}
\usepackage{amssymb}
\usepackage{amsfonts}

\usepackage{graphicx}
\usepackage{epsfig}
\usepackage{amssymb}

\def\pp{{\bf p}}

\def\xx{{\bf x}}
\def\LL{{\cal L}}
\def\Uf{{F}}
\def\Upp{{\bf P}}

\def\Ut{{T}}

\def\DUt{{\partial_{\Ut}}}

\def\Rset{\hbox{{I\kern -0.2em R}}}
\def\rset{\hbox{{\tiny\rm I\kern -0.2em R}}}

\def\QQ{Q}

\begin{document}
\title{Propagation in quantum walks and relativistic diffusions}
\author{Fabrice Debbasch}
\email{fabrice.debbasch@gmail.com}
\affiliation{ UPMC, ERGA-LERMA, UMR 8112, 3, rue Galil\'ee, F-94200 Ivry, France}
\author{Giuseppe Di Molfetta}
\affiliation{ UPMC, ERGA-LERMA, UMR 8112, 3, rue Galil\'ee, F-94200 Ivry, France}
\author{David Espaze}
\affiliation{ UPMC, ERGA-LERMA, UMR 8112, 3, rue Galil\'ee, F-94200 Ivry, France}
\author{Vincent Foulonneau}
\affiliation{ UPMC, ERGA-LERMA, UMR 8112, 3, rue Galil\'ee, F-94200 Ivry, France}
\bibliographystyle{unsrt}

\begin{abstract}
Propagation in quantum walks is revisited by showing that very general 1D discrete-time quantum walks with time- and space-dependent coefficients can be described, at the continuous limit, by Dirac fermions coupled to electromagnetic fields. 
Short-time propagation is also established for relativistic diffusions by presenting new numerical simulations of the Relativistic Ornstein-Uhlenbeck Process. A geometrical generalization of Fick's law is also obtained for this process.
The results suggest that relativistic diffusions may be realistic models of decohering or random quantum walks. Links with general relativity and geometrical flows are also mentioned.
\end{abstract}
\maketitle

\section{Introduction}

Classical random walks are non quantum, non relativistic models of diffusion in which both time and space are discrete. The simplest formal quantum analogues of random walks are the Discrete-Time Quantum Walks (DTQWs), which have been introduced in \cite{ADZ93a} and \cite{Meyer96a}. They can be used to model transport in solids \cite{Bose03a,Aslangul05a,Burg06a, Bose07a}, disordered media \cite{AVG98a,Mourach98a,Wester06a} and even complexes of algae \cite{Engel07a,Collini10a}. They also constitute simple systems where decoherence can be fruitfully explored \cite{var96a,Perets08a} and are also of great importance in quantum information and quantum computing \cite{VA12a}.

Relativistic diffusion models have been developed, not only to deal with relativistic particle transport, but also to serve as templates to build consistent relativistic hydrodynamical theories \cite{I87a} and realistic models of non relativistic transport at bounded velocity \cite{KGVS96a,CSL04a,IMS05a,KG07a,JRKMG08a,CDR08b}. To obtain relativistic diffusion models, one must (i) jettysone discretization and treat space and time as continuous (ii) introduce equations of motion which fix the position of the particle, not only in space-time, but in relativistic phase-space. The simplest model of physical relevance is the Relativistic Ornsetin-Uhlenbeck Process (ROUP), which has been introduced in \cite{DMR97a}. Other processes have been proposed, notably by Franchi-LeJan \cite{FlJ07a} and Dunkel-Haenggi \cite{DH05a}. A unified presentation can be found in \cite{CD07g}. We refer the reader to \cite{DC07b,DH08a} for more information on this rapidly growing field. 

The aim of this article is to present new results addressing propagation in both DTQWs and relativistic diffusions. 
Here is a summary of the material. Analytically, propagation can be ascertained by showing that the dynamics is controlled by a wave equation. 
It is  well-known that at least some DTQWs display propagative features \cite{Strauch06a}. These features are best highlighted by examining the continuous limits of quantum walks. We revisit this problem by considering the most general 1D quantum walks with time-and position-dependent coefficients.
The object which can admit a continuous limit is {\sl not} generically a given discrete-time walk, but rather a 1-jet of discrete-time walks, which regroups all walks which share a common expansion in the time-and length steps as these tend to zero. Only certain jets obeying particular scaling laws admit a continuous limit. We investigate the richest scaling law and prove that the continuous limit of the associated 1-jets is described by a Dirac equation obeyed by a fermion coupled to an electronagnetic field. 

Analytical results on relativistic diffusions are difficult to obtain and exist in the long-time limit only, where a large class of models (including the ROUP) actually behave as classical traditional Galilean diffusions. We have studied the short- and intermediate time behaviour of the ROUP through numerical simulations of the associated transport equation. Our results show that, contrary to all expectations, the ROUP displays propagative behaviour at short times. We also present numerical evidence that the time-evolution of the density profile of the ROUP strongly resembles the time-evolution of a decohering
\cite{Ken07a} or random \cite{Konno05a} quantum walk. We finally show that the ROUP obeys a geometrical generalization of Fick's law which links relativistic diffusions with black hole physics and geometrical flows.

\section{Propagation in quantum walks}

\subsection{Discrete-time quantum walks in $(1 + 1)$ dimensions}

We consider quantum walks defined over discrete time and discrete one dimensional space, driven by
time- and space-dependent quantum coins acting on a two-dimensional Hilbert space $\mathcal H$. 
The walks are defined by the following finite difference equations, valid for all $(j, m) 
\in \mathbb{N}  \times \mathbb{Z}$: 
\begin{equation}
\begin{bmatrix} \psi^{-}_{j+1, m }\\ \psi^{+}_{j+1, m } \end{bmatrix} \  = 
B\left( \theta_{j, m} ,\xi_{j, m} ,\zeta_{j, m}, \alpha_{j,m} \right)
 \begin{bmatrix} \psi^{-}_{j, m+1} \\ \psi^{+}_{j, m-1} \end{bmatrix},
\label{eq:defwalkdiscr}
\end{equation}
where 
\begin{equation}
 B(\theta ,\xi ,\zeta, \alpha) = e^{i \alpha}
\begin{bmatrix}  e^{i\xi} \cos\theta &  e^{i\zeta} \sin\theta\\ - e^{-i\zeta} \sin\theta &  e^{-i\xi} \cos\theta.
 \end{bmatrix}
\label{eq:defB}
\end{equation} 
This operator is in SU(2) only for $\alpha$ = $p \pi$, $p \in \mathbb {Z}$ and $\theta$, $\xi$ and $\zeta$ are then called the three Euler angles of $B$.
The index $j$ labels instants  and the index $m$ labels spatial points.
The wave function $\Psi$ has two components $\psi^- $ and $\psi^+$ which code for the probability amplitudes 
of the particle jumping towards the left or towards the right.
The total probability $\pi_j= \sum_m \left(
\mid \psi^- _{j, m} \mid ^2+ \mid \psi^+ _{j, m}\mid^2 \right)$ is independent of $j$ {\sl i.e.} 
conserved by the walk.
The set of angles $\left\{ \theta_{j, m} ,\xi_{j, m} ,\zeta_{j, m},\alpha_{j,m} (j, m) \in \mathbb{N} 
 \times \mathbb{Z}\right\}$ 
defines the walks and at this stage is arbitrary with the only resriction on $\alpha$ seen above to save 
the unitarity of the propagation.  
To investigate the continuous limit,
we first introduce a time step $\Delta t$ and a space step $\Delta x$.
 We then introduce, for any quantity $a$ appearing in (\ref{eq:defwalkdiscr}), a function $\tilde a$ defined on $\mathbb{R}^+ \times \mathbb{R}$ such that the number $a_{j,m}$ is the value taken by $\tilde a$ at the space-time point $(t_j = j \Delta t, x_m = m \Delta x)$. Equation (\ref{eq:defwalkdiscr}) then reads:
\begin{equation}
\begin{bmatrix} \psi^{-}(t_j+\Delta t, x_m) \\ \psi^{+}(t_j+\Delta t, x_m) \end{bmatrix} \  = 
B
 \begin{bmatrix} \psi^{-}(t_j, x_m+\Delta x) \\ \psi^{+}(t_j, x_m-\Delta x) \end{bmatrix},
\label{eq:defwalk}
\end{equation}
where the tilde's have been dropped on all functions to simplify the notation and the euler angles in the matrix 
are time and space dependent.
We now suppose, that all functions can be chosen at least $C^2$ in both space and time variables for all 
sufficiently small values of $\Delta t$ and $\Delta x$.  
The formal continuous limit is defined as the couple of differential equations obtained from (\ref{eq:defwalk}) 
 by letting both $\Delta t$ and $\Delta x$ tend to zero.  

\subsection{Scaling laws in $(1 + 1)$ dimensions}

Let us now introduce a time-scale $\tau$, a length-scale $\lambda$, an infinitesimal $\epsilon$ and write
\begin{eqnarray}
\Delta t & = & \tau \epsilon \nonumber \\
\Delta x & = & \lambda \epsilon^\delta,
\end{eqnarray}
where $\delta >0$ traces the fact that $\Delta t$ and $\Delta x$ may tend to zero differently.
For the continuous limit to exist, at least formally, the operator $B({\theta ,\xi ,\zeta, \alpha})$  
must also tend to unity as $\epsilon$ tends to zero. This is so because the two column vectors of the left-hand 
side and on the right-hand side of (\ref{eq:defwalk}) both tend to $\Psi(t_j, x_m)$  when $\Delta t$ and $\Delta x$ tend to zero. 
Let us define the scaling laws for the angles by the relations:
\begin{eqnarray}
 \theta(t, x)& = \theta_0 + {\bar \theta}(t, x) \epsilon^\alpha \nonumber \\
 \xi(t, x)& = \xi_0 + {\bar \xi}(t, x) \epsilon^\beta \nonumber \\
 \zeta(t, x)& = \zeta_0 + {\bar \zeta}(t, x) \epsilon^\gamma \\ 
 \alpha(t,x)& = \alpha_0 + {\bar \alpha}(t,x) \epsilon^\eta \nonumber
\end{eqnarray}
where the four exponents $\alpha$, $\beta$, $\gamma$ and $\eta$ are all positive.

Imposing that B tends to unity as $\epsilon$ tends to zero leads to:
\begin{eqnarray}
e^{i(\alpha_0 + \xi_0)} \cos\theta_0 & = & 1 \nonumber \\
 e^{i(\alpha_0  -\xi_0 )} \cos\theta_0 & = & 1 \nonumber \\
 e^{i(\zeta_0+\alpha_0)} \sin\theta_0 & = & 0 
\end{eqnarray}
These can only be satisfied if there is a integer $p \in \mathbb Z$ such that 
$\theta_0 = \alpha_0 = p \pi$ and $\xi_0 = 0$.
Note that there is no constraint on $\zeta_0$ {\sl i.e.} on the limit of the Euler angle $\zeta$ at
$\epsilon = 0$.

The continuous limit can then be investigated by Taylor expanding $\psi^\pm (t, x  \mp \Delta x)$, 
$\psi^\pm (t \pm \Delta t, x)$, $\cos\theta$, $\sin \theta$ and $e^{i \xi}$ in powers of $\epsilon$.  
Possible relationship between the exponents $\alpha$, $\beta$, $\gamma$ and $\eta$ are found by 
examining the lowest order contributions.
One has:
\begin{equation}
\psi^\pm (t \pm \Delta t, x) = \psi^\pm (t, x) + O(\epsilon),
\end{equation}
\begin{equation}
\psi^\pm (t, x \mp \Delta x) = \psi^\pm (t, x) + O(\epsilon^\delta),
\end{equation}
\begin{equation}
e^{i(\alpha + \xi)} \cos\theta = 1 + O(\epsilon^\beta) + O(\epsilon^\eta) + O(\epsilon^{2 \alpha}),
\end{equation}
and
\begin{equation}
e^{i(\alpha + \zeta)} \sin\theta =  O(\epsilon^\alpha) 
\end{equation}
Equation (\ref{eq:defwalk}) then leads to:
\begin{equation}
\begin{bmatrix} \psi^{-}(t_j, x_m) \\ \psi^{+}(t_j, x_m) \end{bmatrix} + O(\epsilon) = 
\begin{bmatrix} \psi^{-}(t_j, x_m) \\ \psi^{+}(t_j, x_m) \end{bmatrix} + \sum_{i=\alpha, \beta, \eta, \delta} O(\epsilon^i) 
\label{eq:basescal}
\end{equation}
Zeroth order contributions cancel out as expected and the remaining terms must balance each other. 
The richest and most interesting case corresponds to $\alpha = \beta = \delta = \eta= 1$ because all contributions 
to (\ref{eq:basescal}) are then of equal importance. This is the scaling law that will be investigated in the remaining of this article.

\subsection{Dirac equation}


Dimensionless time and space coordinates are $T = t/\tau$ and $X = x/\lambda$. 
The natural space-time coordinates to investigate the above scaling law turn out to be the so-called null coordinates $u^-$ and $u^+$, defined in terms of $X$ and $T$ by  
\begin{eqnarray}
u^- & = & \frac{1}{2} \left(T - X\right) \nonumber \\
u^+ & = & \frac{1}{2} \left(T + X \right).
\end{eqnarray}
Partial derivatives with respect to these null coordinates read:
\begin{eqnarray}
\partial_- & = & \partial_{u^-}  = \partial_T -  \partial_X   \nonumber \\
\partial_+ & = & \partial_{u^+} = \partial_T+ \partial_X,
\label{eq:partialuv}
\end{eqnarray}.

The expansion of the discrete equations leads, at first order in $\epsilon$, to the following
continuous equations of motion for $\psi^-$ and $\psi^+$\, :
\begin{equation}
 \partial_-\psi^-  = i (\overline{\alpha} + \overline{\xi} )\psi^- + 
\overline{\theta} e^{+i\zeta}\psi^+ 
\label{eq:Dirac-}
\end{equation}
and 
\begin{equation}
 \partial_+\psi^+  =  i (\overline{\alpha} - \overline{\xi}) \psi^+ - 
\overline{\theta} e^{-  i\zeta}\psi^- 
\label{eq:Dirac+}
\end{equation}

Taken together, these two coupled first-order equations look like a Dirac equation in $(1 + 1)$ dimensions.
 Let us thus recall a few well-known fact about Dirac equations in $2$ space-time dimensions. 
In flat two dimensional space-times, the Clifford algebra can be represented by $2 \times 2$ matrices 
acting on two-component spinors \cite{Davies11}. This algebra admits two independents generators $\gamma^0$ and $\gamma^1$, 
which can be represented by $2 \times 2$ matrices obeying the usual anti-commutation relation:
\begin{equation}
\{\gamma^a, \gamma^b\} = 2 \eta^{ab} {\mathcal I},
\end{equation}
where $\eta$ is the Minkovski metric and $\mathcal I$ is the identity (unity) matrix. 
%
Consider the representation
$\gamma^0 = \sigma_1$ and $\gamma^1 = - \sigma_1 \sigma_3 = i \sigma_2$.
where $\sigma_1$, $\sigma_2$ and $\sigma_3$ are the three Pauli matrices:
\begin{equation}
\sigma_1 = 
\begin{bmatrix}  0 &  1\\ 1 &  0
\end{bmatrix}
\label{eq:defsigma1}
\end{equation}
\begin{equation}
\sigma_2 = 
\begin{bmatrix}  0 &  -i\\ i &  0
 \end{bmatrix}
\label{eq:defsigma2},
\end{equation}
\begin{equation}
\sigma_3 = 
\begin{bmatrix}  1 &  0\\ 0 &  -1
 \end{bmatrix}.
\label{eq:defsigma3}
\end{equation}

The equations (14) and (15) can be recast in the following compact form:
\begin{eqnarray}
 (i\gamma^0 D_0 + i \gamma^1 D_1 - \mathcal{M}) \Psi = 0
\end{eqnarray}
where $D_\mu$ = $\partial_\mu$ - i $A_\mu$, $\partial_0 = \partial_T$, $\partial_1 = \partial_X$, 
$A_0 = \bar \alpha$, $A_1 = - \bar \xi$, 
\begin{equation}
\mathcal{M} = {\bar \theta} \exp\left( - i \mu \sigma_3 \right) = {\mbox{diag}}\left(m^-, m^+\right)
\end{equation}
with
\begin{equation}
m^\pm = {\bar \theta} \exp\left( \pm i \mu\right)
\end{equation}
and
\begin{equation}
\mu = \frac{\pi}{2} + \zeta.
\end{equation}

Equation is a generalized Dirac-equation is $(1 + 1)$ dimensions. 
The coupling between $\Psi$  and the electromagnetic field $(A_0, A_1)$ is standard, 
but the mass term is not: for arbitrary $\zeta$, the masses of the two components $\psi^-$ and $\psi^+$ are 
neither identical nor real, but are complex conjugate to each other. These two masses are equal and real
 if $\exp(2i \mu) = 1$ {\sl i.e.} if there exists and integer $k \in {\mathbb Z}$ such that 
$\zeta = (2k - 1) \pi/2$. 
The common mass $m$ is them equal to $(-1)^k {\bar \theta}$ and its sign thus depends on both $k$ and ${\bar \theta}$.

\section{Propagation in the Relativistic Ornstein-Uhlenbeck Process}

\subsection{Fundamentals}

The ROUP is a stochastic process which describes the diffusion of a special relativistic point mass $m$ in a fluid at
equilibrium with temperature $\theta_e$.
It is completely defined \cite{DMR97a}, in $n$ space dimensions, 
by the equations of motion of the point mass in the rest frame 
of fluid\,:
\begin{eqnarray}
  d\xx&=&\displaystyle\frac{\pp}{m\gamma}dt
  \label{eq:LangevinX}\\
  d\pp&=&\displaystyle-\alpha \frac{\pp}{\gamma}\,dt+\sqrt{2D}\,d{\bf B}_t,
  \label{eq:LangevinP}
\end{eqnarray}
where $\gamma=\sqrt{1+(\frac{\pp}{mc})^2}$ and $\pp^2$ is the squared Euclidean norm of $\pp$.
Equation~(\ref{eq:LangevinX}) is simply the definition of the relativistic $n$-momentum in terms 
of the velocity \cite{LL75a}. Equation~(\ref{eq:LangevinP}) states that the force acting on the particle splits 
into two contributions. The first one is a deterministic friction 
$-\alpha \pp/\gamma$, which forces the $n$-momentum to relax to the vanishing $n$-momentum of
the fluid in which the particle diffuses, and the second one is a $n$-dimensional centered Gaussian white noise. The noise coefficient $D$ is linked to $\alpha$ and $\theta_e$ by the fluctation-dissipation relation $k_B \theta_e = D/\alpha$.

The temperature $\theta_e$ and the mass $m$ define the thermal velocity $v_{\mbox{th}} = (k_B \theta_e/m)^{1/2}$; this velocity, combined with the time-scale $\alpha^{-1}$, defines the length-scale $\lambda_{\mbox{th}} = v_{\mbox{th}} \alpha^{-1}$. We use these characteristic scales to define dimensionless 
time, position and momentum variables $T$, $X$ and $P$ by $T = \alpha t,$ ${\bf X} = {\bf x}/\lambda_{\mbox{th}}$ and ${\bf P} = {\bf p}/(m v_{\mbox{th}})$. The relativistic character of the problem is traced by the quotient $$Q = \frac{c}{v_{\mbox{th}}} = \left( \frac{mc^2}{k_B \theta_e}\right)^{1/2}.$$This parameter is the dimensionless value of the light velocity $c$. The Galilean limit corresponds to $\QQ\rightarrow\infty$ and the so-called ultrarelativistic case corresponds to $Q \rightarrow 0$.

The equations of motion, being stochastic, do not generate a single trajectory from a given initial condition, but rather an infinite set of possible trajectories. The position of the particle in its phase-space $\Rset^{2n} = \{{\bf X}, {\bf P} \}$ is thus represented, for a given initial condition and value of $Q$, by a probability density $F_Q(T, {\bf X}, {\bf P})$ with respect to $d^nX d^n P$. This density 
obeys the forward Kolmogorov equation
\begin{equation}
  \DUt \Uf_Q + \nabla_{\bf X} \cdot \left( \frac{\bf P}{\Gamma_\QQ(\Upp)} \Uf_Q\right)= \LL_\QQ\,\Uf_Q,
  \label{eq:DedimFP}
\end{equation}
where 
\begin{equation}
\LL_\QQ\,\Uf_Q  = \nabla_\Upp\cdot\Bigl(\frac{\Upp}{\Gamma_\QQ(\Upp)} \,  \Uf\Bigr)+\Delta_\Upp\Uf_Q
\label{eq:defLQ}
\end{equation}
and $\Gamma_\QQ(\Upp)=\sqrt{1+(\Upp/\QQ)^2}$ is the Lorentz factor.

The J\"uttner distribution \cite{J11a} describing a relativistic thermal equilibrium at temperature $\theta_e$ reads simply\, :
\begin{equation}
  \Uf^\star_\QQ(\Upp)=A\,\exp{\Bigl(-\QQ^2\Gamma_\QQ(\Upp)\Bigr)},
  \label{eq:DedimJuttner}
\end{equation}
where $A$ is the normalization factor; this distribution obeys ${\mathcal L}_Q \Uf^\star_Q = 0$.

The density of the process $N_Q(T, {\bf X})$ and the particle current density ${\bf J}_Q(T, {\bf X})$ at time $T$ and position ${\bf X}$ are defined by:
\begin{equation}
N_Q(T, {\bf X}) = \int_{\rset^n} F_Q(t, X, P) d^nP
\end{equation}
and 
\begin{equation}
{\bf J}_Q (T, {\bf X}) = \int_{\rset^n} \frac{\bf P}{\Gamma_Q({\bf P})} F_Q(T, {\bf X}, {\bf P}) d^nP.
\end{equation}
They obey the continuity relation 
$\partial_T N_Q + \partial_{\bf X} \cdot {\bf J}_Q = 0$.
Fick's law posits that the current density and density gradient are proportional to each other with constant coefficient. This simple assumption is not valid for the ROUP. The generalized Fick's law obeyed by the ROUP is presented in Section ***  below.

 The spatial Fourier transform ${\hat F}_Q$ of the phase-space density $F_Q$ is defined by
\begin{equation}
{\hat F}_Q (T, {\bf K}, {\bf P}) = \frac{1}{\sqrt{2 \pi}} \, \int_{\rset^n} F_Q(T, {\bf X}, {\bf P}) \exp
\left( i {\bf K} \cdot {\bf X}\right) d^nX
\end{equation}
and obeys:
\begin{equation}
  \DUt {\hat F}_Q + i {\bf K} \cdot \left( \frac{\bf P}{\Gamma_\QQ(\Upp)}{\hat F}_Q \right)= \LL_\QQ\,{\hat F}_Q.
  \label{eq:DedimhatF}
\end{equation}
The Fourier tranform ${\hat N}_Q$ is defined in a similar manner and coincides with the integral of ${\hat F}_Q$ over $P$. Suppose a diffusing particle is put into the fluid at time $T = 0$ with initial position ${\bf X} = 0$ and 
initial temperature $\theta = \theta_e$. The initial condition for equation (\ref{eq:DedimFP}) is then
\begin{equation}
{F}_Q (T= 0, {\bf X}, {\bf P}) =  \delta( {\bf X}) \Uf^\star_\QQ(\Upp),
\label{eq:Fhat0}
\end{equation}
or, in Fourier space,
\begin{equation}
{\hat F}_Q (T = 0, {\bf K}, {\bf P}) =  \frac{1}{\sqrt{2 \pi}} \Uf^\star_\QQ(\Upp).
\label{eq:Fhat0}
\end{equation}
This initial condition fully determines, for each value of $Q$, the value of density $F_Q$ at all times and phase-space positions.
Equation (\ref{eq:DedimhatF}) does not involve any derivation with respect to ${\bf K}$, which can thus be viewed as a simple parameter (as opposed to a fully-fledged variable with respect to which derivations are performed). Numerical simulations have been carried out by solving (\ref{eq:DedimhatF}).
The function $F_Q$ has been recovered by Fast Fourier Transform and the spatial density $N_Q$ as well as the current $J_Q$ have been obtained from $F_Q$ by direct numerical quadrature. All simulations have been performed with Mathematica 8. Various integration methods and limit conditions in $P$ have been used, to ensure the robustness of numerical results.

\subsection{The density profile of the ROUP}


The spatial density $N_Q(T, X)$ at time $T$ and point $X$ vanishes for $\mid X \mid > QT$ (remember $Q$ is the dimensionless value of the light velocity $c$). To obtain the best visual representation of the early time evolution of the density profile, we introduce the rescaled position variable $\xi = X/QT$ and the rescaled density $\nu_Q (T, \xi) = N_Q(T, QT \xi)/(QT)$, which is normalized to unity against the measure $d\xi$.
Figures 1 and 2 present typical plots of $\nu_Q$ against $\xi$ for different values of $T$ and $Q$. At early times, the maximum of the density profile is not situated at $\xi = 0$ {\sl i.e.} at the starting point of the diffusion, but rather at $\mid \xi \mid = \xi_Q$ where $\xi_Q$ is time-independent and approximately equal to 0.948. This corresponds to $X_Q (T) \approx 0.948\,  QT$ {\sl i.e.} $x_Q(t) \approx 0.948\,  ct$. This means that the diffusion, at early times, mostly propagates at velocity ${\tilde c} \approx 0.95\,  c$.
In time, a secondary maximum appears at the origin point $\xi = 0$. This secondary maximum grows and finally becomes much higher than the peaks at $\pm \xi_Q(T)$. The density profile thus gets closer and closer to the standard Gaussian $G(T, X)$ predicted by Fick's law
\cite{DR98a, AF07a}.

\subsection{A heuristic analytical argument supporting the observed short-time propagation}


At early times, the Fourier transform ${\hat F}_Q$ of the phase-space density is {\sl a priori} close to its initial value $F_Q^*/\sqrt{2 \pi}$.
As indicated above, ${\mathcal L}_Q F_Q^* = 0$.
Thus, at early times, the Fourier transform ${\hat F}_Q$ obeys approximately 
\begin{equation}
 \DUt {\hat F} + i {K}  \left( \frac{P}{\Gamma_\QQ(P)}{\hat F} \right) = 0,
\end{equation}  
 which can be integrated easily into:
\begin{equation}
{\hat F}_Q(T, K, P) \approx \frac{1}{\sqrt{2 \pi}} F^*_Q(P) \exp \left( - i \, \frac{K P}{\Gamma_Q(P)} t\right).
\end{equation}
Carrying out the inverse Fourier transform and the integration over $P$ leads to 
\begin{equation}
{N}_Q(T, X) \approx \frac{1}{2\pi T} \left(\gamma_Q\left(\frac{X}{T}\right)\right)^3  \exp \left( - Q^2 \gamma_Q\left( \frac{X}{T}\right) \right)
\end{equation}
where
$\gamma_Q(V) = (1 - V^2/Q^2)^{-1/2}$ 
is the Lorentz factor associated to the velocity $V$.
This function of $X/T$ strongly resembles the early time density profile of the ROUP, with the characteristic `valley'  shape around the origin and maxima at $\mid X/T \mid  = 2 {\sqrt 2}Q/3 \approx 0.943\;Q$, remarkably close to the numerically observed maxima situated at  $0.948\,  QT$ (see the above first section). It also tends very rapidly to zero as $X/T$ tends towards $\pm Q$ because the Lorentz factor tends to infinity at these points, and the exponential thus tends to zero.

\subsection{Generalized Fick's law}

One might wonder if the propagative behaviour of the ROUP is compatible with any generalization of Fick's law and what this generalization, if it exists, looks like. The ROUP does obey a generalization of Fick's law and this generalization is best understood in geometrical terms.
Each Riemanian metric $g$ defined over an $n$-dimensional physical space defines in a canonical manner a Brownian motion on this space. This result is well-known for time-indepedent metrics \cite{Ok98a} but has been extended recently  \cite{CD07d,CD07h,ACT08a,CD10a,C10a} to time-dependent ones. For this Brownian motion, the link between the density $N$ (with respect to the $g$-independent measure $dX$), the current density $J$ and the metric $g$ reads
\begin{equation}
J(T, X) =\frac{1}{\sqrt{g(T, X)}}\,  \partial_X \left( \frac{N(T, X)}{\sqrt{g(T, X)}} \right).
\label{eq:GenFick}
\end{equation}
where $n = 1$ has been assumed. This last equation is a clear generalization of Fick's law.
Changing point of view, this equation 
can also be viewed as a differential equation to be solved for the metric $g$ at given density density $N$ and current $J$. Let $g_Q$ be a metric thus associated to $N_Q$ and $J_Q$. 
This metric is best obtained from $N_Q$ and $J_Q$ by rewriting 
(\ref{eq:GenFick}) in terms of $h_Q = g_Q^{-1}$, which leads to\, :
\begin{equation}
\frac{N_Q}{2}\,  \partial_X h_Q + h_Q\,  \partial_X N_Q = -J_Q.
\end{equation}
We choose as solution $h_Q(T, X) = I_Q(T, X)/N_Q^2(T, X)$ with
\begin{equation}
I_Q(T, X) = -2 \int_{-QT}^X N_Q(T, Y) J_Q(T, Y) dY.
\end{equation}
The standard, Galilean Ornstein-Uhlenbeck process corresponds to $Q = + \infty$; as shown in \cite{DR08a}, $J_{\infty} = - \chi(T) \partial_X N_\infty$, so that $h_\infty(T, X) = \chi(T)$, which is flat and tends (as it should) to $1$, {\sl i.e.} the time-independent Euclidean metric, as $T$ tends to infinity.

Typical results are displayed in Figure 3. The metric is nearly flat at the centre of the interval $(-QT, QT)$ but grows to infinity near $\mid X \mid \sim QT$. Consider for example a point with coordinate $X$ and a point with coordinate $X + \Delta X$, $\delta X \ll X$. As far as the diffusion is concerned, the effective distance between these two points at time $T$ is $g_Q(T, X) \Delta X$. This distance grows to infinity for any finite $\Delta X$ when $X$ approaches $\pm QT$. Thus, the distance that a particle needs to travel to get closer to $QT$ by the amount $\Delta X$ tends to infinity as the particle approaches $\pm QT$.  This prevents the particle from ever crossing $X = \pm QT$ {\sl i.e.} from being transported faster than light.

Let us add that the metric $g_Q$ can be used to construct the space-time metric ${\tilde g}_Q$, defined by $ds^2 = dT^2 - g_Q(T, X) dX^2$, that ${\tilde g}_Q$ is conformal to $ds^2 = dT^2/\sqrt {g_Q(T, X)} - \sqrt {g_Q(T, X)} dX^2$, and that this last metric admits an horizon at $X = cT$.

We finally remark that the apparently simpler generalization of Fick's law $J = - D(T, X) \partial_X N$ is ruled out by numerical simulations (data not shown) because the current $J_Q$, in the short-time propagative regime, does not vanish at the two maxima of $N_Q$. A similar simple proportionality between $J_Q$ and $N_Q$ in Fourier space is also ruled out, for similar reasons.

\section{Discussion}

Quantum walks and relativistic diffusions are  generalizations of classical random walks. We have presented new evidence, both analytical and numerical, that quantum walks and relativistic diffusions, contrary to classical diffusions, display propagative behaviour. At the continuous limit, the very general unitary quantum walks we have considered in this article display propagation at all times while the ROUP only displays propagation at short times. It is quite remarkable that 
the density profile of the ROUP seems to follow the same time-evolution as the density profiles of non-unitary, decohering quantum walks \cite{Ken07a} and of random quantum walks \cite {Konno05a} (which do not admit continuous limits). We thus suggest that decohering and random quantum walks may be adequately modelled by relativistic diffusions. If that is indeed true, observing quantum walks decohere would constitute laboratory experiments of relativistic transport.

As mentionned earlier, the generalized Fick's law presented in Section III.D has an obvious connection with black hole physics. It also defines a natural geometrical flow on the unit $n$-dimensional ball. If the above connection between relativistic diffusions and quantum walks is true, this means that decohering quantum transport also connects with general relativity and geometrical flows.

\paragraph{Acknowledgements:} Part of this work was funded by the ANR Grant 09-BLAN-0364-01.


\def\cprime{$'$} \def\cprime{$'$} \def\cprime{$'$} \def\cprime{$'$}
  \def\cprime{$'$}

\newpage

\begin{figure}
    \centerline{
    \includegraphics[width=6cm,clip=true]{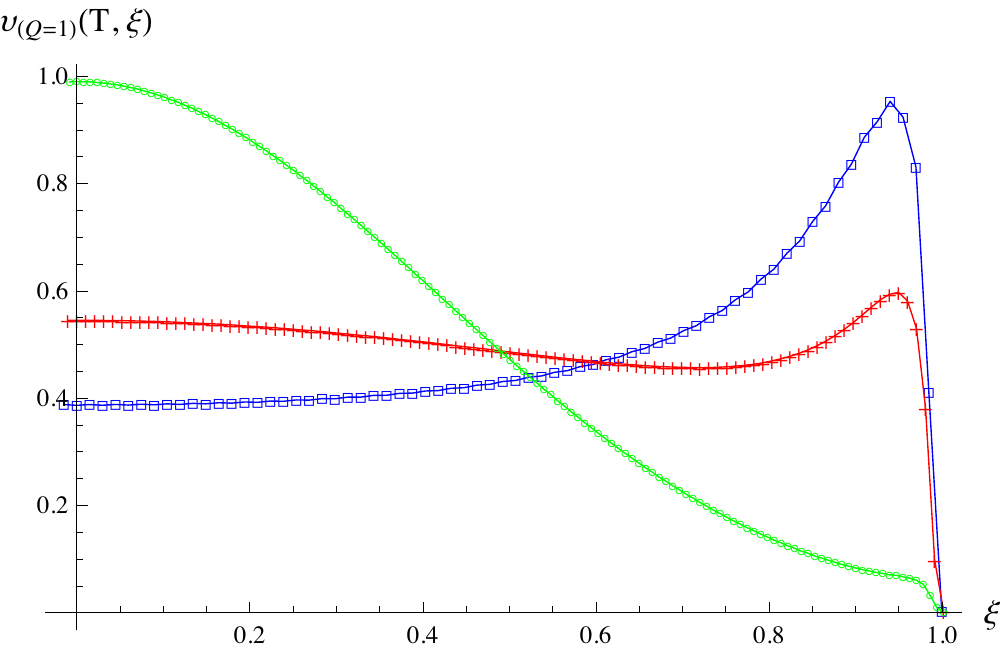}
}
\caption{{\bf Density profile of the ROUP at fixed $Q$} Rescaled density-profile $\nu_Q$ against rescaled position $\xi = x/(ct)$ for $Q = 1$
and $T = 0.5$ (blue squares), $T = 2$ (red triangles) and $T = 10$ (green circles).}
\end{figure}

\begin{figure}
    \centerline{
     \includegraphics[width=6cm,clip=true]{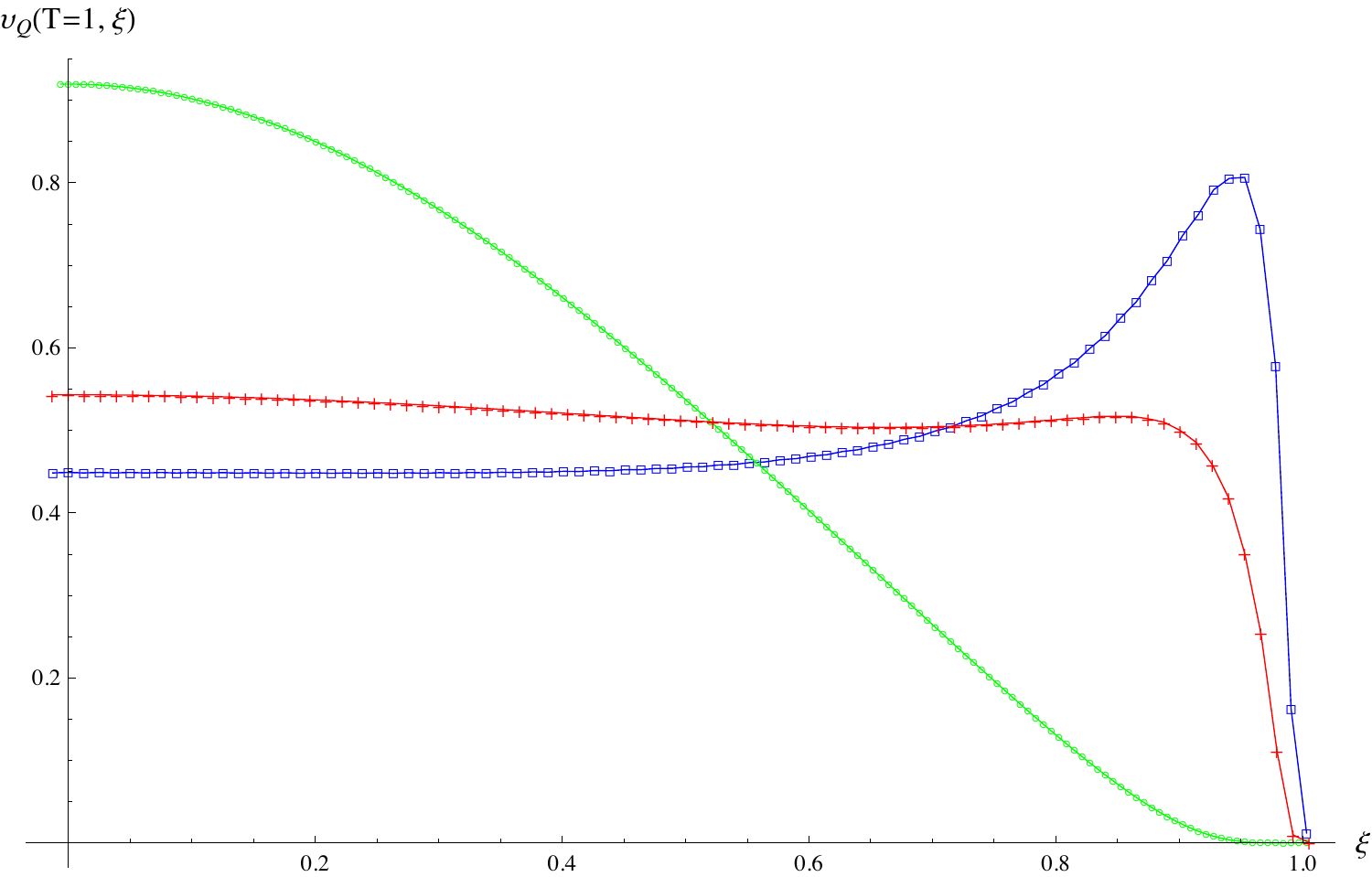}
 }
  \caption{{\bf Density profile of the ROUP at fixed $T$} Rescaled density-profile $\nu_Q$ against rescaled position $\xi = x/(ct)$ for $T = 1$
and $Q = 1$ (blue squares), $Q = 1.2$ (red triangles) and $Q = 2$ (green circles).}
  \label{fig:Dens_Prof2}
\end{figure}

\nopagebreak

\begin{figure}
    \centerline{
    \includegraphics[width=6cm,clip=true]{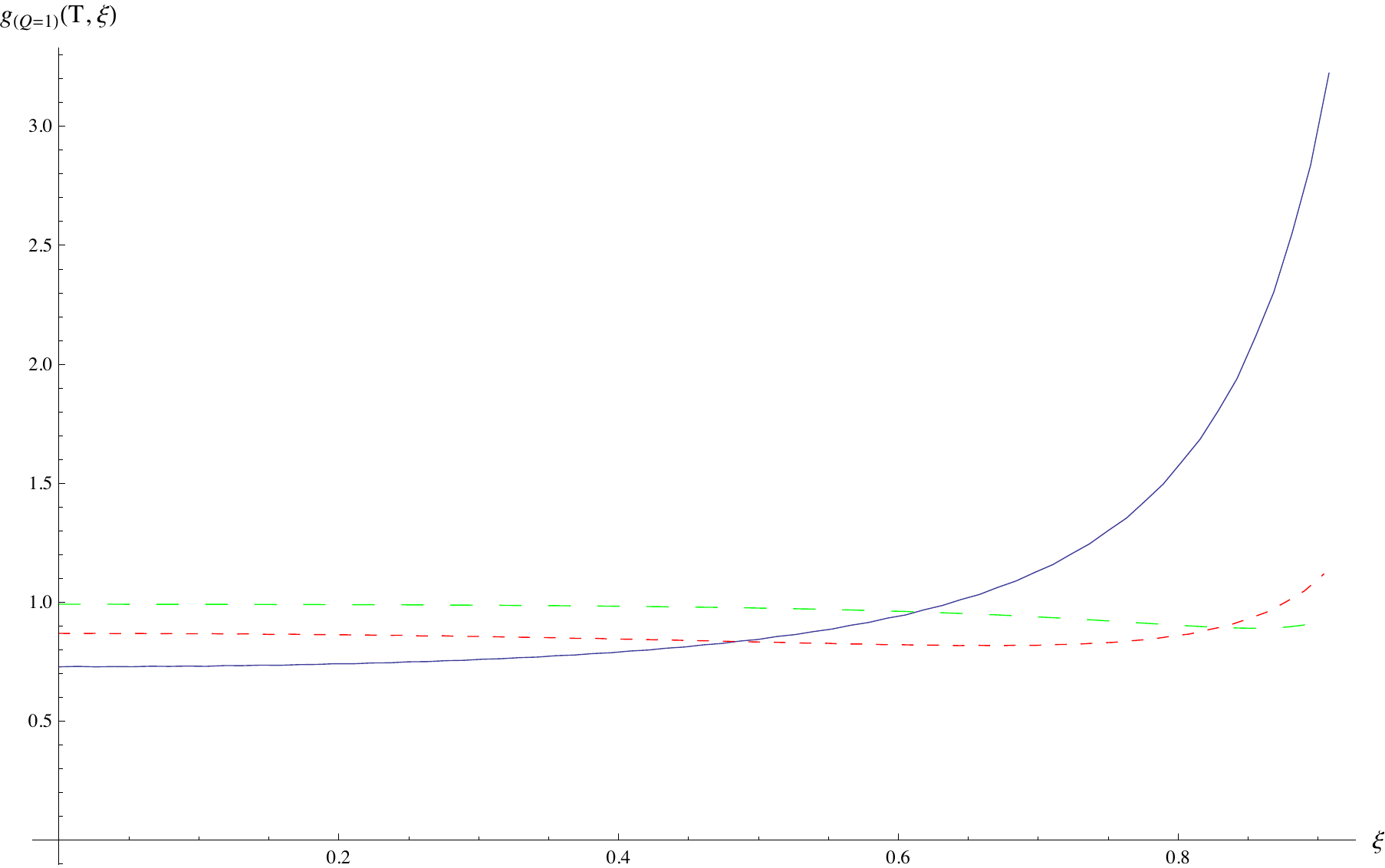}
}
  \caption{{\bf Diffusion metric appearing in the generalized Fick's law.} Metric $g_Q$ plotted against the rescaled variable $\xi = x/(ct)$ for $Q = 1$ and $T = 1$ (blue curve), $T = 4$ (red small dashes curve), $T = 10$ (green large dashes curve)}
\label{fig:Metric}
\end{figure}

\end{document}